\documentclass[utf8]{frontiersSCNS}
\usepackage{url,hyperref,lineno,microtype,subcaption}
\usepackage[onehalfspacing]{setspace}
\def\keyFont{\fontsize{8}{11}\helveticabold }
\def\firstAuthorLast{Reichhardt {et~al.}}
\def\Authors{C. J. O. Reichhardt$^*$ and C. Reichhardt}
\def\Address{Theoretical Division and Center for Nonlinear Studies,
  Los Alamos National Laboratory, Los Alamos, New Mexico 87545, USA}
\def\corrAuthor{C. J. O. Reichhardt}

\def\corrEmail{cjrx@lanl.gov}

\begin{document}
\onecolumn
\firstpage{1}
\title[Fluctuations and Pinning For Individually Manipulated Skyrmions]{Fluctuations and Pinning For Individually Manipulated Skyrmions}

\author[\firstAuthorLast ]{\Authors}
\address{}
\correspondance{}

\extraAuth{}

\maketitle
 
\begin{abstract}

\section{}  
We numerically examine the dynamics of individually dragged skyrmions interacting simultaneously with an array of other skyrmions and quenched disorder. For drives just above depinning, we observe a broad band noise signal with a $1/f$ characteristic, while at higher drives, narrow band or white noise appears. Even in the absence of quenched disorder, the threshold force that must be applied to translate the driven skyrmion is finite due to elastic interactions with other skyrmions. The depinning threshold increases as the strength of the quenched disorder is raised. Above the depinning force, the skyrmion moves faster in the presence of quenched disorder than in a disorder-free system since the pinning sites prevent other skyrmions from being dragged along with the driven skyrmion. For strong pinning, we find a stick-slip motion of the driven skyrmion which produces a telegraph noise signature. The depinning threshold increases monotonically with skyrmion density in the absence of quenched disorder, but when pinning is present, the depinning threshold changes nonmonotonically with skyrmion density and there are reentrant pinned phases due to a competition between pinning induced by the quenched disorder and that produced by the elastic interactions of the skyrmion lattice.

\tiny
 \keyFont{ \section{Keywords:} skyrmion, dynamic phases, broad band noise, telegraph noise, depinning}
  
\end{abstract}

\section{Introduction}

Magnetic skyrmions in chiral magnets are particle-like textures that form a
triangular lattice [\cite{Muhlbauer09,Yu10}]
and can be set into motion under various
types of drives [\cite{Nagaosa13,Iwasaki13,Schulz12,Woo16}].
Moving skyrmions can interact with each other other as well
as with impurities or quenched disorder in the sample
[\cite{Nagaosa13,Reichhardt21}].
One consequence 
is the presence of a finite depinning threshold
or critical driving force needed to set the skyrmions in motion.
Depinning thresholds have been observed that span
several orders of magnitude depending on
the properties of the materials [\cite{Schulz12,Woo16,Reichhardt21}].
Another interesting aspect of skyrmions is that their motion is
strongly influenced by gyroscopic effects 
or the Magnus force. This force appears in addition to the dissipative
effects that can arise from Gilbert damping and other sources.
In the absence of quenched disorder, the Magnus force
causes a driven skyrmion to move at a finite angle
known as the skyrmion Hall angle $\theta_{sk}$ with respect to the driving force,
where the value of $\theta_{sk}$
is proportional to the ratio of the Magnus to the damping
forces
[\cite{Nagaosa13,Reichhardt21,EverschorSitte14,Reichhardt15,Jiang17,Litzius17}].
When quenched disorder is present, $\theta_{sk}$ becomes
velocity or drive dependent, starting from a zero value at low drives
and gradually increasing with increasing
velocity until it saturates at high drives to a value close to the intrinsic or
disorder free Hall angle
[\cite{Reichhardt15,Jiang17,Litzius17,Reichhardt19,Juge19,Zeissler20}].
Skyrmion depinning and motion can also be probed using
the time series of the skyrmion velocity.
Both numerical
and experimental studies have shown that near the depinning transition,
the skyrmion motion is disordered
and the system exhibits large noise fluctuations with
broad band or $1/f^\alpha$ features,
while at higher drives there is a crossover to
white noise or even a narrow band or periodic noise signal
[\cite{Diaz17,Sato19,Sato20}].
The onset
of narrow band noise is
an indication that the skyrmions have formed
a periodic lattice structure.
Similar transitions between broad and narrow band noise as a function of
drive have also been observed
for the depinning and sliding dynamics
of vortices in type-II superconductors [\cite{Marley95,Olson98a}],
driven charge density waves [\cite{Gruner88}],
and other driven assemblies of particles
moving over random quenched disorder [\cite{Reichhardt17}].

Interest in skyrmion dynamics and pinning is driven in part by the prospect of
using skyrmions in a variety of applications
[\cite{Fert17,Luo21}].
Many of these applications require the manipulation of individual
skyrmions or the interaction of skyrmions 
with a disordered landscape, so understanding the motion and fluctuations
of individually manipulated skyrmions would be
a valuable step in this direction.
There have been numerous studies
of methods to manipulate or drag individual particles
with and without quenched disorder
which focused on the
velocity and fluctuations of the manipulated particle.
Examples include driving single colloids through assemblies of other
colloids
[\cite{Hastings03,Habdas04,Zia18,Dullens11,Gazuz09}],
as well as
measuring the changes of the effective viscosity on the driven particle
as the system goes through glass [\cite{Hastings03,Habdas04,Gazuz09}],
melting [\cite{Zia18,Dullens11}]
or jamming transitions [\cite{Candelier10,Reichhardt10}]. 
Other studies have explored how
the depinning threshold changes in a clean system as
the system parameters are varied [\cite{Habdas04,Reichhardt08,Gruber20}],
as well as the effect of quenched disorder on individually manipulated
superconducting vortices and magnetic textures
[\cite{Straver08,Auslaender09,Veshchunov16,Kremen16,Ma18}].
It is also possible to examine changes in the fluctuations as a function
of drive
while the density of the surrounding medium or the
coupling to quenched disorder is changed 
[\cite{Candelier10,Reichhardt10,Ma18,Illien18}].
In experiments on skyrmion systems,
aspects of the pinning landscape have been examined by
moving individual skyrmions with local tips
[\cite{Hanneken16,Holl20}].
It is also possible to drag
individual skyrmions with optical traps [\cite{Wang20b}] or
by other means [\cite{Reichhardt21a}]
and to examine
the motion of the skyrmions within the traps as well as changes
in the velocity and skyrmion Hall angle as function of driving force.
Most of the extensive numerical and experimental studies of
the dynamics of individually dragged particles have
focused on bulk properties such as the average velocity or
effective drag coefficients,
and there is little work examining how
the time series, noise fluctuations, or depinning threshold
of a single probe particle would change when quenched disorder is present.
This is of particular interest
for skyrmions,
since one could expect different fluctuations to appear
in the damping dominated regime compared
to the strong Magnus or gyroscopic dominated regime.

In this work we introduce quenched disorder to the system in order to
expand on our previous study [\cite{Reichhardt21a}] of driving
individual skyrmions through an assembly of other skyrmions.
We specifically focus on the time series of the velocity fluctuations,
noise power spectra, effective drag, and changes in
the depinning threshold while varying
the ratio of the Magnus force to the damping.
For strong damping, we generally find
enhanced
narrow band noise signals.
We show that although quenched disorder
can increase the depinning threshold,
it can also decrease the drag experienced by the driven particle
and reduce the amount of 
broad band noise.
In the absence of quenched disorder,
the depinning threshold monotonically increases with increasing 
system density [\cite{Reichhardt21a}],
but we find that when quenched disorder is present,
the depinning becomes strongly nonmonotonic due
to the
competition between the pinning from the quenched disorder and the
pinning from elastic interactions with the surrounding medium. 
This can also be viewed as an interplay between
pinning [\cite{Reichhardt17}] and jamming [\cite{Reichhardt14}] behaviors.

\section{Simulation and System} 

We consider a modified Thiele equation
[\cite{Thiele73,Lin13,Brown18}] or particle-based approach
in which a single skyrmion is driven though a two-dimensional assembly
of other skyrmions and a quenched disorder landscape.
The initial skyrmion positions are obtained using simulated annealing,
so that in the absence of quenched disorder,
the skyrmions form a triangular lattice.
The equation of motion of the driven skyrmion is given by
\begin{equation}
\alpha_d {\bf v}_i - \alpha_m \hat{\bf z} \times {\bf v}_i  = {\bf F}_i^{ss} + {\bf F}_i^{p} + {\bf F}^{D}_i.
\end{equation}
Here, the instantaneous velocity is  ${\bf v}_i=d{\bf r}_i/dt$,
${\bf r}_i$ is the position of skyrmion $i$, and $\alpha_d$ is the damping
coefficient arising from dissipative processes.
The gyroscopic or Magnus force,
given by the second term on the left hand side,
is of magnitude $\alpha_m$ and
causes the skyrmions to move in the direction perpendicular to
the net applied force.
The repulsive skyrmion interaction force has the form
[\cite{Lin13}]
${\bf F}_i^{ss} = \sum_{j = 1}^{N_s} K_1(r_{ij}) \hat{\bf r}_{ij}$,
where $r_{ij} = |{\bf r}_i - {\bf r}_j|$,
$\hat{\bf r}_{ij} = ({\bf r}_i - {\bf r}_j)/r_{ij} $,
and $K_1$ is the
modified Bessel function which decays exponentially for large $r$.
Within the system are $N_{p}$ non-overlapping randomly placed pinning sites which
are modeled as
parabolic traps with a maximum range of $r_{p}$ that produce a pinning force
given by 
${\bf F}_i^{p} = \sum_{k = 1}^{N_p} (F_p/r_p) ({\bf r}_i - {\bf r}_k^{(p)}) \Theta( r_p - |{\bf r}_i - {\bf r}_k^{(p)}| )\hat{\bf r}_{ik}^{(p)}$, where
$F_{p}$
is the maximum pinning force and $\Theta$ is the Heaviside step
function.
The driving force ${\bf F}^{D}_i = F_{D} \hat{\bf x}$
is applied only to a single skyrmion.
Under this driving force, in the absence of pinning or collisions
with other skyrmions the skyrmion would move with
an intrinsic skyrmion Hall angle of
$\theta^{\rm int}_{sk} = \arctan(\alpha_{m}/\alpha_{d})$.
We measure the net skyrmion velocity
${\bf V}=N_s^{-1}\sum_{i=1}^{N_s}{\bf v}_i$ and its time-averaged components
parallel, $\langle V_{||}\rangle$, and perpendicular,
$\langle V_{\perp}\rangle$,
to the driving force, which is applied along the $x$ direction.
The measured skyrmion Hall angle is
$\theta_{sk}  = \arctan(\langle V_{\perp}\rangle/\langle V_{||}\rangle)$.
The sample is of size $L \times L$ with $L=36$, and in most of this
work we 
consider $N_s=648$, giving a skyrmion density of $n_s=N_s/L^2=0.5$,
and $N_p=388$, giving a pinning site density of $n_p=N_p/L^2=0.3$.

In previous work, we considered a similar model
containing no pinning [\cite{Reichhardt21a}], where
a finite critical depinning force $F_c$ for motion of the driven
skyrmion arises due to
elastic interactions with the background skyrmions.
There is also a higher second depinning force $F^{tr}_{c}$
at which 
the driven skyrmion begins to move transverse to the driving direction,
producing a finite skyrmion Hall angle.
$\theta_{sk}$ increases
with increasing drive until,
for high drives, it reaches
a value close to the intrinsic value $\theta_{sk}^{\rm int}$.
This is similar to the behavior found for an assembly of skyrmions 
driven over random  disorder
[\cite{Reichhardt15,Jiang17,Litzius17,Reichhardt19,Juge19,Zeissler20,Diaz17}].
For a fixed drive, the net velocity of the driven skyrmion
can actually increase with increasing system density
due to the Magnus-induced velocity boost effect,
whereas in the overdamped limit, the
velocity decreases monotonically with increasing density due to
enhanced damping from the
increased frequency of collisions with background skyrmions
[\cite{Reichhardt21a}].
In the present work, we study the effects of adding quenched disorder, and
we measure
time dependent velocity fluctuations,
velocity overshoots,
and the depinning threshold.
The time series can be characterized by the
power spectrum
\begin{equation}
S(\omega) = \left|\int V(t)\exp(-i \omega)dt\right|^{2}
\end{equation}
Power spectra
provide
a variety of information on the dynamical properties of the
system [\cite{Weissman88}]
and have been used extensively to characterize
depinning phenomena
[\cite{Marley95,Olson98a,Gruner88,Reichhardt17,Bullard08,Reichhardt16}]. 
In this work we focus specifically on the fluctuations
of the velocity component in the direction of drive.

\section{Results}

\begin{figure}[h!]
\begin{center}
\includegraphics[width=10cm]{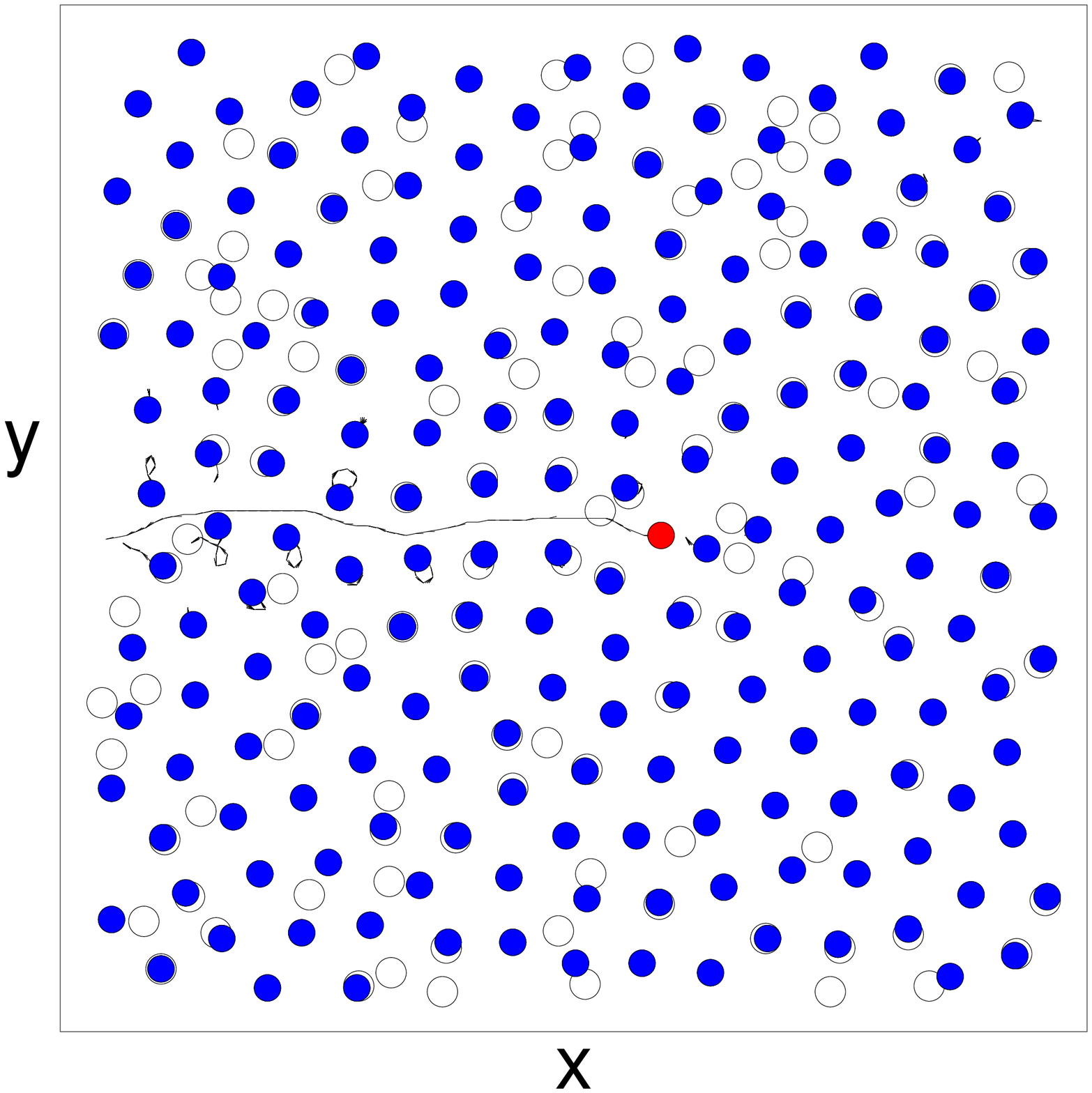}
\end{center}
\caption{ 
An image of a subsection of the system
in which a single skyrmion (red) is driven though an assembly of other
skyrmions (blue) in the presence of quenched disorder, generated by
randomly placed nonoverlapping local trapping sites (open circles). 
Black lines indicate the skyrmion trajectories.
The driven skyrmion generates motion of the surrounding skyrmions
as it passes through the system.
}
\label{fig:1}
\end{figure}

In Fig.~\ref{fig:1} we illustrate a subsection of the system
containing a single skyrmion driven though a
background of other skyrmions (blue dots) and pinning sites (open circles).
The skyrmion trajectories indicate that
the driven skyrmion creates a distortion in the surrounding
medium as it moves through the system.

\begin{figure}[h!]
\begin{center}
  \includegraphics[width=10cm]{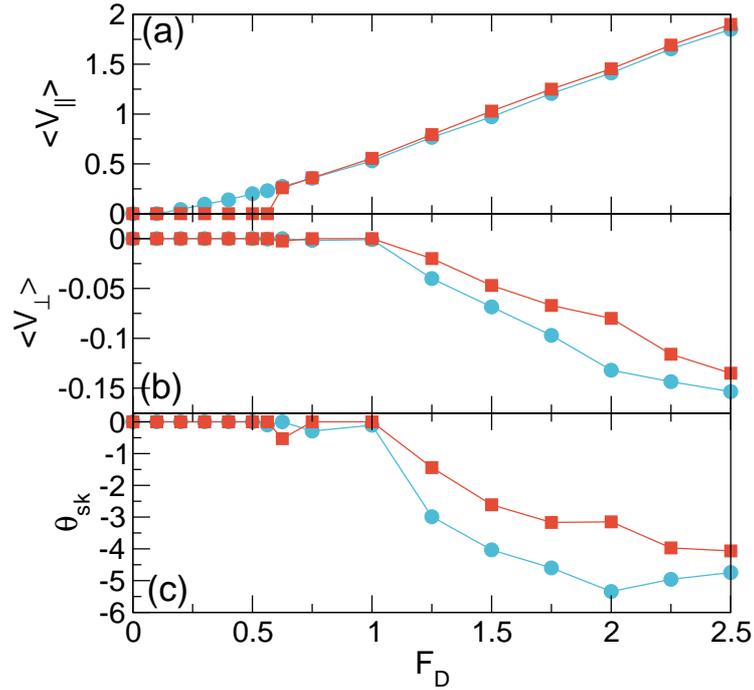}
\end{center}
\caption{
a) The velocity $\langle V_{||}\rangle$ parallel to the driving direction versus drive $F_D$ for the system in Fig.~\ref{fig:1} with
$n_{s} = 0.5$,  $\alpha_{m} = 0.1$, $\alpha_{d} = 0.995$,
and $\theta^{\rm int}_{sk} = -5.74^\circ$.
Blue circles are for a system with no pinning, $n_p=0$, while red squares are
for a system
with $n_{p} = 0.3$ and $F_{p} = 0.3$.
(b) The corresponding velocity in the perpendicular direction
$\langle V_{\rm \perp}\rangle$ versus $F_D$.
(c) The corresponding measured skyrmion Hall angle $\theta_{sk}$ versus $F_D$.
}
\label{fig:2}
\end{figure}

In Fig.~\ref{fig:2}(a) we plot the average velocity parallel to the
drive, $\langle V_{||}\rangle$, versus $F_{D}$ for
the system in Fig.~\ref{fig:1} with $n_{s} = 0.5$,
$\alpha_{m} = 0.1$, and $\alpha_{d} = 0.995$. Here, we employ the constraint
$\alpha^2_{d} + \alpha^2_{m} = 1.0$ [\cite{Reichhardt21a}], and
the intrinsic skyrmion Hall angle is $\theta^{\rm int}_{sk} = -5.74^\circ$.
In the absence of quenched disorder, where $n_p=0$,
a depinning threshold appears near $F_{c}^{np} = 0.1$.
For
$0.1 < F_{D} \leq 1.0$, the skyrmion is moving but,
as indicated in Figs.~\ref{fig:2}(b) and (c),
$\langle V_{\perp}\rangle = 0$ and thus $\theta_{sk} = 0^\circ$.
For $F_{D} > 1.0$, $\langle V_{\perp}\rangle$ becomes finite and $\theta_{sk}$
begins to grow in magnitude with increasing $F_D$ until it
saturates near $\theta_{sk} = -4.0^\circ$.
In a sample containing pinning with $n_p=0.3$ and $F_p=0.3$,
where the ratio of skyrmions to pinning sites is $5:3$,
the depinning threshold appears
at $F_{c} = 0.565$.
This value is higher than what would be observed
in the single skyrmion limit, where
$F_{c}^{ss} = F_{p} = 0.3$,
indicating that the skyrmion-skyrmion interactions are playing an
important role in the depinning process.
It is also higher
than the sum $F_{c}^{ss} + F^{np}_{c}=0.4$ of the single skyrmion
and pin-free thresholds,
showing that the skyrmions at the pinning sites produce an enhanced
pinning effect on the driven skyrmion.
In the sample with quenched disorder,
$\langle V_{||}\rangle$ is finite for $0.565 < F_{D} \leq 1.0$
but the corresponding $\langle V_{\perp}\rangle=0$,
while
for $F_{D} > 1.0$,
both $\langle V_{\perp}\rangle$ and $\theta_{sk}$ increase in magnitude
with increasing $F_D$.
In the regime $F_{D} > 1.0$,
$\langle V_{||}\rangle$ for the system containing pinning
is higher than that found in the system without pinning.
This is a result of the fact that in the clean system the driven skyrmion
pushes some of the background skyrmions along with it, creating an enhanced
drag which reduces $\langle V_{||}\rangle$, whereas
when pinning is present, the surrounding skyrmions are trapped by the pinning
sites and cannot be entrained to move along with the driven skyrmion.
The reverse trend appears in $\langle V_{\perp}\rangle$,
where both the perpendicular velocity and the skyrmion Hall angle
are smaller in magnitude when pinning is present than for the system without
pinning.

\begin{figure}[h!]
\begin{center}
  \includegraphics[width=10cm]{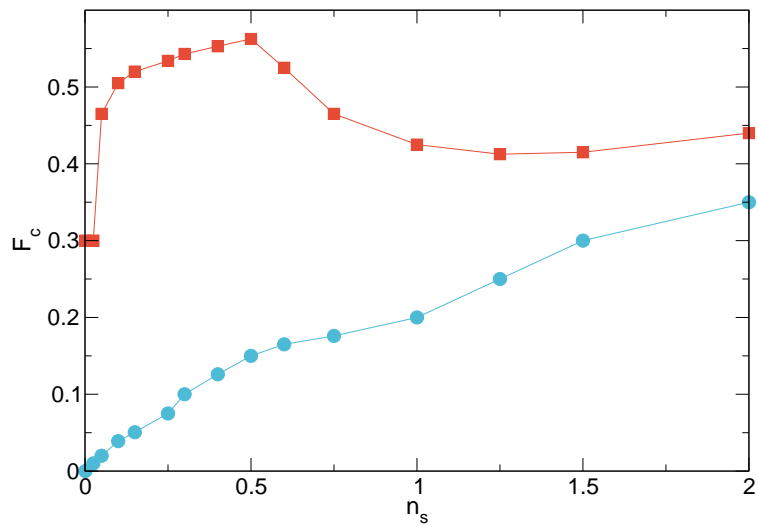}
\end{center}
\caption{ 
The depinning force $F_{c}$ versus skyrmion density $n_s$
for the systems in Fig.~\ref{fig:2} with $\alpha_m=0.1$, $\alpha_d=0.995$,
and $\theta_{sk}^{\rm int}=-5.74^\circ$.
Blue circles are for a pin-free system with $n_p=0$ while red squares are
for a system with $n_p=0.3$ and $F_p=0.3$.
With no quenched disorder, $F_c$ increases monotonically
with $n_s$.
}
\label{fig:3}
\end{figure}

In Fig.~\ref{fig:3} we plot the depinning force $F_{c}$
versus skyrmion density $n_{s}$ for the systems in Fig.~\ref{fig:2} with
and without pinning.
In the absence of pinning,
$F_{c}$ starts from zero and increases monotonically
with increasing $n_{s}$ as it becomes 
more difficult to push the skyrmion through the system.
When pinning is present,
at low $n_{s}$ the driven skyrmion interacts only with the pinning sites,
giving $F_{c} = F_{p}$;
however, once the density increases enough for the driven skyrmion to
interact with both pinning sites and other skyrmions,
$F_{c}$ sharply increases and reaches a maximum value near $n_{s} = 0.5$.
The
maximum depinning force $F^{\rm max}_c$ should be approximately equal to
the force needed to depin the
driven skyrmion from a pinning site plus the force required to
push the skyrmion directly in front of the driven skyrmion out of
a pinning site, $F^{\rm max}_c=2F_p$, which is close to the value we observe.
Up to $n_s=0.5$, the driven skyrmion can always find an empty pinning site
to occupy. If the pinning were periodic, all pins would be filled once
$n_s=n_p$, but since the pinning is randomly placed, some pins remain empty
and available even when $n_s$ is somewhat larger than $n_p$.
Once $n_{s} > 0.5$, $F_{c}$ decreases
with increasing $n_{s}$ because the driven skyrmion is no longer able to
find an available pinning site and is instead pinned only by the repulsion
from the neighboring pinned skyrmions. This interstitial pinning is weaker
than the direct pinning, and as $n_s$ increases above $n_s=0.5$,
a larger and larger number of 
interstitial skyrmions appear in the sample, decreasing the fraction of
directly pinned skyrmions and leading to the decrease in $F_c$.
There are, however, still a nonzero fraction of pinned skyrmions, so
$F_c$ remains well above the value found in the sample without pinning
sites.
At large $n_{s}$,
$F_{c}$ begins to increase with increasing density, in line with 
the trend found for the sample with no quenched pinning,
where the
interactions with an increasing number of unpinned skyrmions
makes it more difficult for the driven skyrmion to move through the system.
As $n_{s}$ is increased beyond the range shown in Fig.~\ref{fig:3},
we expect that the curves for the pinned and unpinned samples will
approach each other as the fraction of directly pinned skyrmions
becomes smaller and smaller.

\begin{figure}[h!]
\begin{center}
  \includegraphics[width=10cm]{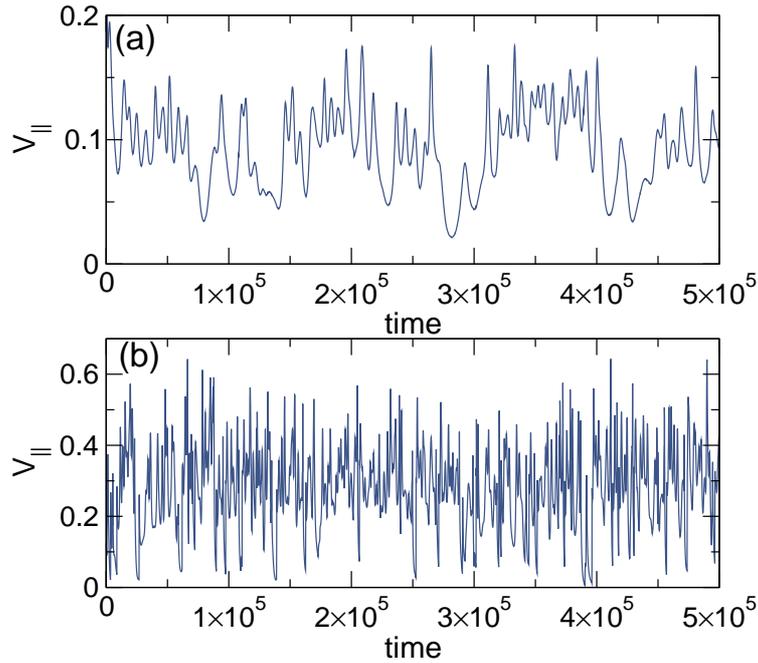}
\end{center}
\caption{ 
The time series of the velocity $V_{||}$ parallel to the drive
for the system in Fig.~\ref{fig:2} with $n_{s} = 0.5$, $\alpha_m=0.1$,
$\alpha_d=0.995$, and $\theta_{sk}^{\rm int}=-5.74^\circ$.
(a) The disorder-free $n_p=0$ system
at $F_{D} = 0.3$ just above depinning,
showing periodic oscillations and longer time plastic events.
(b) The quenched disorder system with $n_p=0.3$ and $F_p=0.3$ 
at $F_{D} = 0.625$ just above depinning, showing less correlated motion.
}
\label{fig:4}
\end{figure}

In Fig.~\ref{fig:4}(a) we plot the time series of the
parallel velocity $V_{||}$ for the system in
Fig.~\ref{fig:2} at $n_{s} = 0.5$ with no quenched disorder for
$F_D  = 0.3$, just above depinning.
A series of short-period oscillations appear which correspond
to elastic interactions in which the driven skyrmion moves past a background
skyrmion without generating plastic displacements of the background skyrmions.
There are also infrequent larger signals in the form of sharp velocity dips
that are correlated with the creation of a plastic distortion or exchange
of neighbors among the background skyrmions due to the passage of the
driven skyrmion.
In Fig.~\ref{fig:4}(b) we show the time series of $V_{||}$ for
the system with quenched disorder
at $F_{D} = 0.625$,
just above the depinning threshold. Here, the motion is much more disordered,
with strong short time velocity oscillations.
These are produced by the motion of the driven skyrmion over the
background pinning sites. The overall structure of the background skyrmions
is disordered, destroying the periodic component of motion
found in the unpinned system.

We next examine the power spectra $S(\omega)$ of
time series such as those shown in Fig.~\ref{fig:4}
for different drives for the systems in Fig.~\ref{fig:2}.
Generically, power spectra can take several forms
including $1/f^\alpha$, where
$\alpha = 0$ indicates white noise with little or no correlation,
$\alpha = 2$ is 
Brownian noise,
and $\alpha = 1$ or pink noise can
appear when large scale collective
events occur [\cite{Weissman88}].
Noise signatures that are periodic produce narrow band
signals with peaks at specific frequencies.
It is also possible to have combinations
in which the signal is periodic on one time scale
but has random fluctuations on longer time scales.
For assemblies of particles under an applied drive
that exhibit plastic depinning, the power spectrum is typically
of $1/f^{\alpha}$ type with
$\alpha$ ranging from $\alpha=1.3$ to $\alpha=2.0$.
A single particle moving
over an uncorrelated random landscape
typically shows a white noise spectrum, while
motion over a periodic substrate produces
narrow band noise features [\cite{Reichhardt17}].

\begin{figure}[h!]
\begin{center}
  \includegraphics[width=10cm]{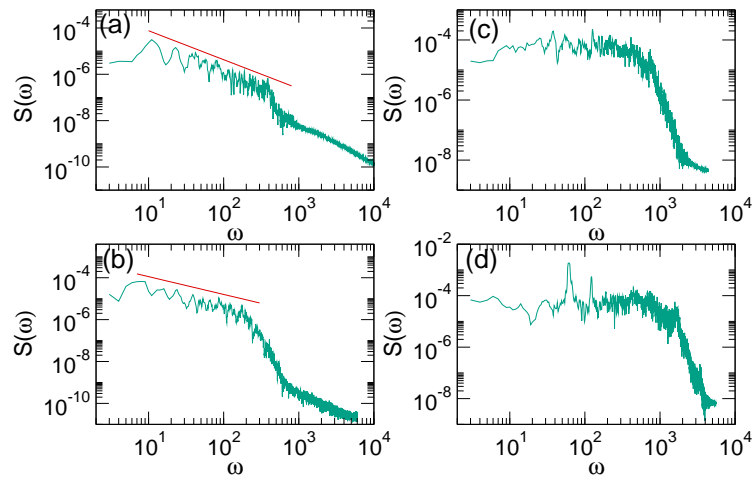}
\end{center}
\caption{(a)
The power spectra $S(\omega)$ for the system in Fig.~\ref{fig:2} with
$n_s=0.5$, $\alpha_m=0.1$, $\alpha_d=0.995$, $\theta_{sk}^{\rm int}=-5.74^\circ$,
and no quenched disorder ($n_p=0$).
(a) At $F_{D} = 0.2$ there is a $1/f^{\alpha}$ signature, where the straight line indicates $\alpha = 1.25$, along with a series of peaks
corresponding to the oscillatory portion
of the motion due to the periodicity of the skyrmion
lattice. (b) At $F_{D} = 0.3$, the red line indicates $\alpha = 0.85$.
(c) At $F_{D} = 1.0$ the signal is white noise with $\alpha = 0$.
(d) At $F_{D} = 1.5$ there is a narrow band noise signal.
}
\label{fig:5}
\end{figure}

In Fig.~\ref{fig:5}(a) we plot $S(\omega)$
for the disorder-free system with $n_p=0$ from Fig.~\ref{fig:2}
at $F_{D} = 0.2$, just above the depinning threshold.
At low frequencies we find
a series of oscillations or a narrow band noise feature.  
These periodic velocity oscillations
correspond to the driven skyrmion speeding up and
slowing down as it moves
through the roughly triangular lattice
formed by the surrounding skyrmions. The
driven skyrmion occasionally generates dislocations
or topological defects in the background lattice,
so the motion is not strictly periodic
but exhibits a combination of periodic motion
with intermittent large bursts.
This intermittent signal is what gives the spectrum an overall
$1/f^{\alpha}$ shape, as indicated by the red line which is
a fit with $\alpha = 1.25$.
The noise power drops at higher frequencies,
which are correlated with the small rotations caused by the Magnus
force as the driven skyrmion generates plastic events.
In Fig.~\ref{fig:5}(b) we show the velocity spectrum in the disorder-free
sample at
$F_{D} = 0.3$ for the time series illustrated in
Fig.~\ref{fig:4}(a). The overall shape of the spectrum is similar to that
found at $F_D=0.2$ in Fig.~\ref{fig:5}(a),
but the low frequency oscillations are
reduced since more plastic events are occurring in the background
skyrmion lattice.
A power law fit with $\alpha = 0.85$ appears as a straight line
in Fig.~\ref{fig:5}(b).
In overdamped
driven systems with quenched disorder,
the power law exponent is observed to decrease with increasing drive
until it reaches a white noise
state with $\alpha = 0$, and a narrow band noise 
signature appears at high drives [\cite{Marley95,Olson98a,Reichhardt17}].
In Fig.~\ref{fig:5}(c) at $F_{D} = 1.0$, the response at lower frequencies
has become a white noise spectrum with
$\alpha = 0$,
while at slightly higher frequencies there is
the beginning of a narrow band noise peak.
At $F_D=1.5$ in Fig.~\ref{fig:5}(d),
strong narrow band peaks appear in the spectrum.
The narrow band noise arises once the driven
skyrmion is moving fast enough that it
no longer has time to generate dislocations or other defects in
the surrounding lattice, making the system appear more like
a single particle moving over a triangular lattice
and creating few to no distortions.
For high drives, the same narrow band noise signal
appears but the peaks shift to higher frequencies as the
driven skyrmion moves faster.

\begin{figure}[h!]
\begin{center}
  \includegraphics[width=10cm]{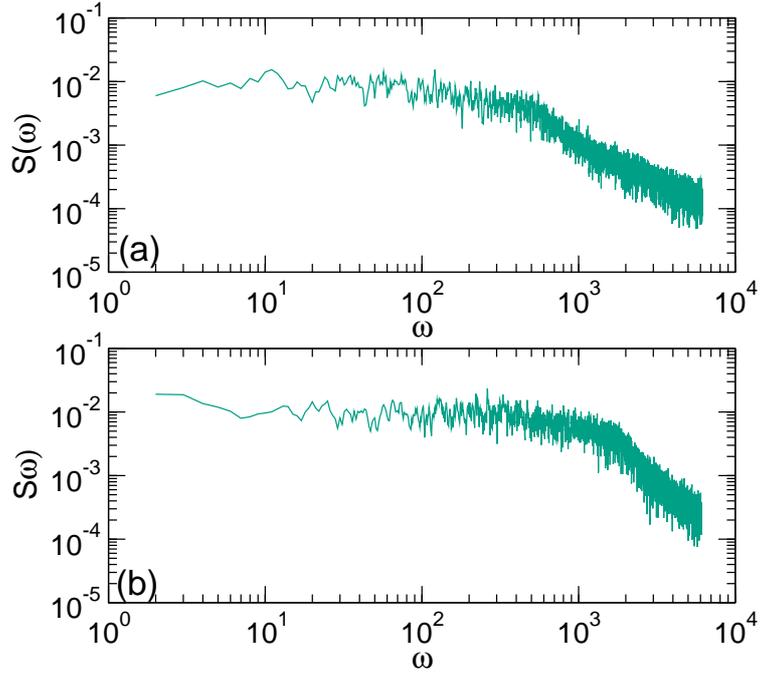}
\end{center}
\caption{
The power spectra $S(\omega)$ for the system in Fig.~\ref{fig:2}
with
quenched disorder at
$n_s=0.5$, $\alpha_m=0.1$, $\alpha_d=0.995$, $\theta_{sk}^{\rm int}=-5.74^\circ$, $n_p=0.3$, and $F_p=0.3$.
(a) At $F_{D} = 0.625$,
the noise signal is close to white with $\alpha = 0$.
(b) A similar spectrum appears at $F_{D} = 1.5$.
The high frequency shoulder
above which a
$1/f^{2}$ signature appears
shifts to higher drives as $F_D$ increases.
}
\label{fig:6}
\end{figure}

In Fig.~\ref{fig:6}(a) we plot $S(\omega)$ for the pinned
system with $n_p=0.3$ and $F_p=0.3$ from Fig.~\ref{fig:4}(b)
at $F_{D} = 0.625$,
just above depinning.
At low frequencies, the power spectrum is nearly white with $\alpha = 0$, 
while the noise power drops as $1/f^2$ at
high frequencies.
Unlike the pin-free system,
strong low frequency oscillations are absent
because the lattice
structure of the surrounding skyrmions is disordered by the pinning sites. 
We find no $1/f$ noise, in part due to the reduced mobility of
the skyrmions trapped at pinning sites, which reduces the amount of plastic
events which can occur.
In the absence of pinning,
the driven skyrmion can more readily
create exchanges of neighbors in the background skyrmions,
generating longer
range distortions in the system
and creating
more correlated fluctuations in the driven skyrmion velocity.
In Fig.~\ref{fig:6}(b) we plot $S(\omega)$ for the same system at
a higher drive of $F_{D} = 1.5$, where again similar
white noise appears at low frequencies, while
the transition from white noise 
to $1/f^2$ noise has shifted to higher frequency.
Unlike the disorder-free sample, here we find
no narrow band signal
since the surrounding
skyrmions are trapped in disordered positions by the pinning. 
The addition of quenched disorder might be expected to
increase the appearance
of $1/f$ noise due to the greater disorder in the system;
however, in this case, the quenched disorder suppresses the
plastic events responsible for the broad band noise signature.
In a globally driven assembly of particles,
the drive itself can induce plastic events
[\cite{Reichhardt17}].
This
implies that the fluctuations 
of a single probe particle
driven over quenched disorder
are expected to differ significantly from the noise signatures found
in bulk driven systems.
The spectra in Fig.~\ref{fig:6} have a shape called
Lorentzian,
$S(f) = A/(\omega_{0}^2 + \omega^2)$,
which is also found for shot noise.
[\cite{Weissman88}]. 
In our case, $\omega_{0}$
corresponds to
the average time between
collisions of the driven skyrmion
with pinning sites, and it shifts to
higher frequencies as the drive increases.

\begin{figure}[h!]
\begin{center}
  \includegraphics[width=10cm]{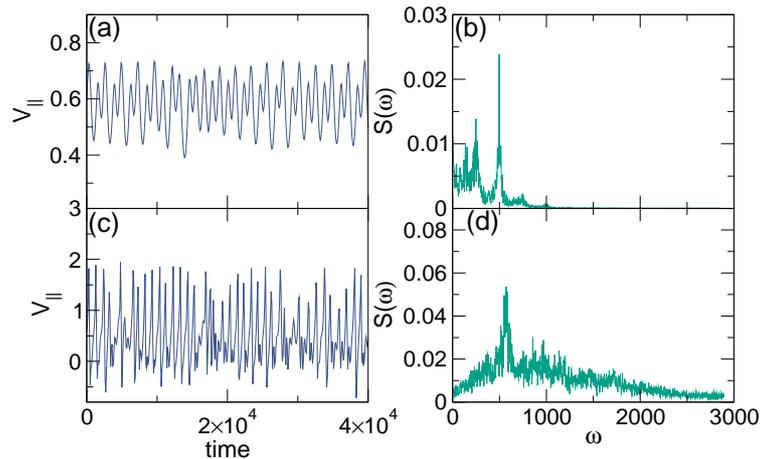}
\end{center}
\caption{
(a) The time series of the velocity $V_{||}$
parallel to the drive
at $n_{s} = 0.5$ and $F_{D} = 1.0$
for an overdamped system with $\alpha_{m}/\alpha_{d} = 0$
and no quenched disorder.
(b) The corresponding power spectrum $S(\omega)$ contains
narrow band peaks.
(c) The time series for the same system but with
$\alpha_{m}/\alpha_{d} = 9.95$, where the fluctuations are enhanced.
(d) The corresponding power spectrum $S(\omega)$ has
a reduced amount of narrow band noise.
}
\label{fig:7}
\end{figure}

\begin{figure}[h!]
\begin{center}
  \includegraphics[width=10cm]{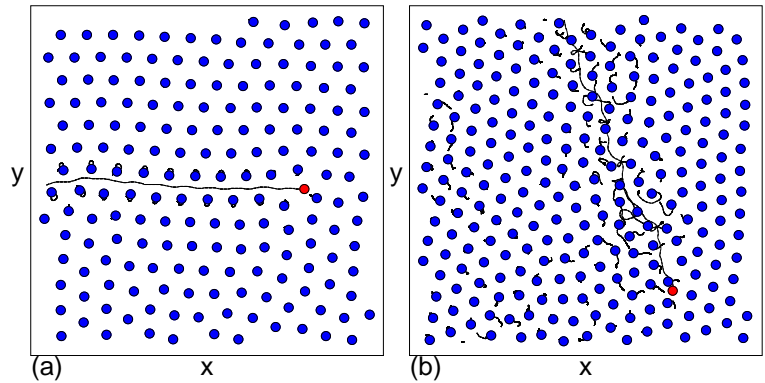}
\end{center}
\caption{
An image of a subsection of the system showing the driven skyrmion (red),
background skyrmions (blue), and skyrmion trajectories (black lines)
for the samples from 
Fig.~\ref{fig:7} with $n_s=0.5$ at $F_D=1.0$.
(a) For the overdamped system with $\alpha_m/\alpha_d=0$ from
Fig.~\ref{fig:7}(a,b), the background skyrmions experience elastic distortions
but there are no plastic events.
(b) For the Magnus dominated system with $\alpha_m/\alpha_d=9.95$ from
Fig.~\ref{fig:7}(c,d),
the driven skyrmion moves at an angle due to the increased Magnus force,
creating significant distortions in the background skyrmion lattice.
}
\label{fig:8}
\end{figure}

We next consider the influence of the Magnus force
on the noise fluctuations of the
driven skyrmion.
In Fig.~\ref{fig:7}(a) we plot
the time series of $V_{||}$
at $F_D=1.0$
for a system without quenched disorder
in the completely overdamped limit
of $\alpha_{m} = 0.0$ and $\alpha_{d} = 1.0$.
For the equivalent drive in a sample with $\alpha_m=0.1$ and
$\alpha_d=0.995$, Fig.~\ref{fig:5}(c) shows that white noise is present;
however, for the overdamped system,
Fig.~\ref{fig:7}(b) indicates that a strong narrow band noise signal
appears.
In the image in Fig.~\ref{fig:8}(a),
the driven skyrmion moves through the lattice of other skyrmions
without creating
plastic distortions. In general,
we find that in the overdamped limit and in the absence of pinning,
a strong narrow band noise signal appears as the driven skyrmion
moves elastically through an ordered skyrmion lattice, as shown
in the linear-linear plot of $S(\omega)$ in Fig.~\ref{fig:7}(b).
In Fig.~\ref{fig:7}(c) we plot the time series
of $V_{||}$ for the same system in the Magnus dominated regime with
$\alpha_{m}/\alpha_{d} = 9.95$
and $\theta^{\rm int}_{sk} = 84.26^\circ$.
Here, a combination of periodic motion and
plastic events occur, producing the 
much smaller narrow band noise signal shown
in Fig.~\ref{fig:7}(d).
In the corresponding skyrmion trajectories illustrated in
Fig.~\ref{fig:8}(b),
the skyrmion moves at an angle to the driving direction due to the
Magnus force,
and there are significant distortions of the surrounding skyrmion
lattice.
This additional motion is generated by
the increase in spiraling behavior produced by the Magnus force.
In previous
studies of bulk driven skyrmions
moving over quenched disorder, it was shown that an increase in the
Magnus force
caused a reduction in the narrow band noise signal [\cite{Diaz17}].

\begin{figure}[h!]
\begin{center}
  \includegraphics[width=10cm]{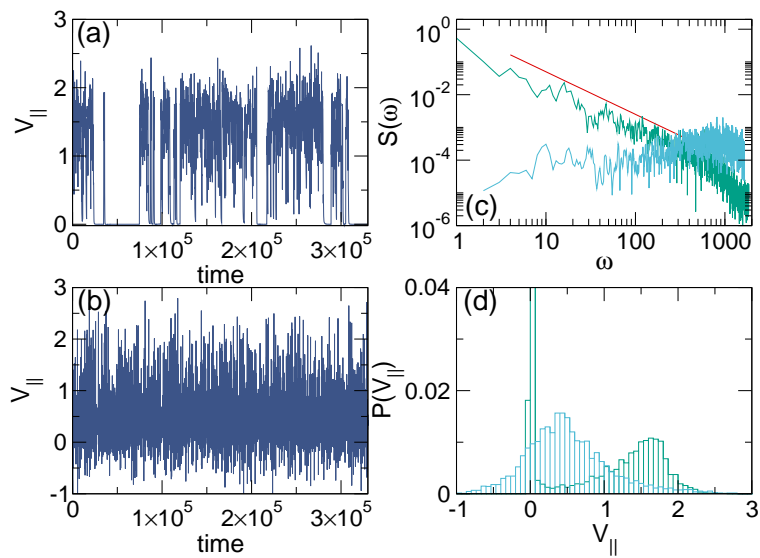}
\end{center}
\caption{
Samples with  
$n_{s} = 0.5$, $n_{p} = 0.3$, $F_{p} = 2.0$, and $F_{D} = 1.6$.
(a) Time series of $V_{||}$
for a system with  $\alpha_{m}/\alpha_{d}  = 0.1$.
(b) Time series of $V_{||}$ for 
a system with
$\alpha_{m}/\alpha_{d}  = 9.95$.
(c) The power spectra $S(\omega)$ for the $\alpha_m/\alpha_d=0.1$ system (green)
showing a power law fit to $1/f^\alpha$ with $\alpha=1.3$ (red line), and
$S(\omega)$ for the $\alpha_m/\alpha_d=9.95$ system (blue).
(d) Distribution $P(V_{||})$ of velocities in the direction parallel to the drive
for the $\alpha_m/\alpha_d=0.1$ system (green), where a bimodal shape
appears, and
for the $\alpha_m/\alpha_d=9.95$ system (blue), where the distribution
is unimodal.
}
\label{fig:9}
\end{figure}

We next consider the effect of the pinning strength
on the dynamics.
In Fig.~\ref{fig:9}(a) we plot the
time series of $V_{||}$ for a system with $\alpha_{m}/\alpha_{d}  = 0.1$,
$n_{s} = 0.5$, $n_{p} = 0.3$, $F_{p} = 2.0$, and
$F_{D} = 1.6$.
Here the driven skyrmion experiences
a combination of sliding and nearly pinned motion,
where at certain points the skyrmion is temporarily trapped
by a combination of the pinning and the skyrmion-skyrmion interactions.
As the surrounding skyrmions
relax, the driven
skyrmion can
jump out of the pinning site where it has become trapped,
leading to another pulse
of motion. This stick-slip or telegraph type motion only
occurs just above the critical driving force
when the pinning force is sufficiently strong,
while for higher drives the motion becomes continuous.
In Fig.~\ref{fig:9}(b) we show
the time series of $V_{||}$
for the same system at $\alpha_{m}/\alpha_{d} = 9.95$,
where the stick-slip or telegraph motion
is lost. 
We note that the value of
$\langle V_{||}\rangle$ in the
Magnus dominated $\alpha_m/\alpha_d=9.95$ system
is smaller than that found in the overdamped $\alpha_m/\alpha_d=0.1$ system
since the increasing Magnus force rotates more of the velocity
into the direction perpendicular to the drive;
however, a similar continuous flow is observed both parallel and perpendicular
to the drive in the Magnus dominated system.
The loss of the stick-slip motion is due to the
increasing spiraling motion of both the driven and background skyrmions.
In Fig.~\ref{fig:9}(c) we plot the
power spectra corresponding to the time series in Figs.~\ref{fig:9}(a,b).
The stick-slip motion of the $\alpha_m/\alpha_d=0.1$ system
produces
a $1/f^{\alpha}$ signature in $S(\omega)$
with
$\alpha = 1.3$.
For the $\alpha_{m}/\alpha_{d} = 9.95$
sample, $S(\omega)$
is much flatter, indicating reduced correlations in the fluctuations,
and also has increased noise power at high frequencies.
The enhanced high frequency noise
results from the
fast spiraling motion of both the driven and the background
skyrmions when they are inside pinning sites.
The detection of enhanced
high frequency noise could thus
provide an indication
that strong pinning effects or strong Magnus forces are present.
In Fig.~\ref{fig:9}(d) we plot the distribution $P(V_{||})$ of instantaneous
velocity for the samples in Figs.~\ref{fig:9}(a,b).
When $\alpha_{m}/\alpha_{d}  = 0.1$,
$P(V_{||})$ is bimodal
with a large peak near $V_{||} = 0$ and a smaller peak near
$V = 1.6$, corresponding to the value of the driving force.
There is no gap of zero weight in $P(V_{||})$ separating
these two peaks.
When $\alpha_{m}/\alpha_{d} = 9.95$, $P(V_{||})$ has only a single
peak at intermediate velocities. Additionally, there is significant
weight 
at negative velocities, which were not present in the strongly damped
sample.
The negative velocities arise when
the skyrmions move in circular orbits due to the
Magnus force
and spend a portion of the orbit moving
in the direction opposite to the driving force. 

\begin{figure}[h!]
\begin{center}
  \includegraphics[width=10cm]{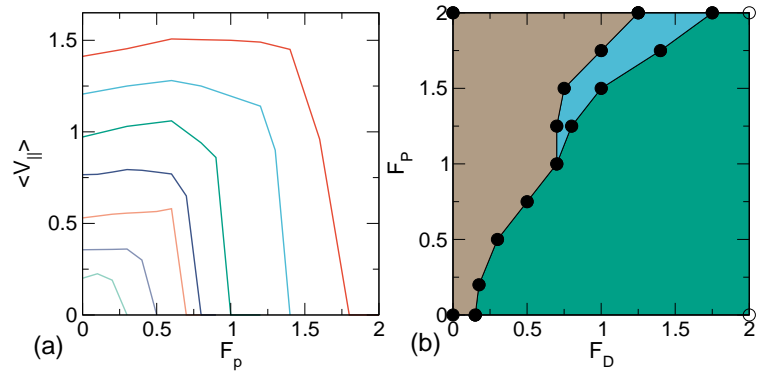}
\end{center}
\caption{
(a) $\langle V_{||}\rangle$ versus pinning strength $F_p$ for
the system in Fig.~\ref{fig:9}(a) with $n_s=0.5$, $n_p=0.3$,
and $\alpha_m/\alpha_d=0.1$ at
$F_D=2.0$, 1.75, 1.5, 1.25, 1.0, 0.7, and 0.5, from top to bottom.
(b) Dynamic phase diagram for the same system as a function of
$F_p$ versus $F_D$. Green: continuous flow regime; blue: stick-slip motion;
brown: pinned.
}
\label{fig:10}
\end{figure}

In Fig.~\ref{fig:10}(a) we plot
the average velocity $\langle V_{||}\rangle$ versus
pinning strength $F_{p}$ for the system in
Fig.~\ref{fig:9}(a) with $\alpha_{m}/\alpha_{d}  = 0.1$,
$n_{s} = 0.5$, and $n_{p} = 0.3$ at 
$F_{D} = 2.0$, 1.75, 1.5, 1.25, 1.0, 0.75, and $0.5$.
The pinning force at which $\langle V_{||}\rangle$ reaches zero,
indicating the formation of a pinned state, increases
as $F_{D}$ increases.
Generally there is also a range of low $F_p$
over which $\langle V_{||}\rangle$
increases with increasing $F_p$.
This is due to a reduction in the drag on the driven skyrmion as the background
skyrmions become more firmly trapped in the pinning sites, similar to what
was illustrated in Fig.~\ref{fig:2}.
Stick-slip motion appears in the regime where there is a sharp
downturn in $\langle V_{||}\rangle$,
and is associated with a bimodal velocity distribution of the type
shown in
Fig.~\ref{fig:9}(d).
A plot of $\langle V_{||}\rangle$ versus $F_p$
for the
$\alpha_{m}/\alpha_{d}  = 9.95$ system (not shown)
reveals a similar trend, except that the pinning transitions shift to
larger values of $F_p$.
Using the features in Fig.~\ref{fig:10}(a) combined with
the velocity distributions, we
construct a dynamic phase diagram for the $\alpha_m/\alpha_d=0.1$ system
as a function of $F_p$ versus $F_D$, illustrated in Fig.~\ref{fig:10}(b).
We observe continuous flow, stick slip motion, and pinned regimes,
with stick-slip motion occurring only when $F_{p} > 0.75$.
In general, for increasing Magnus force,
the window of stick-slip motion decreases in size.

\begin{figure}[h!]
\begin{center}
  \includegraphics[width=10cm]{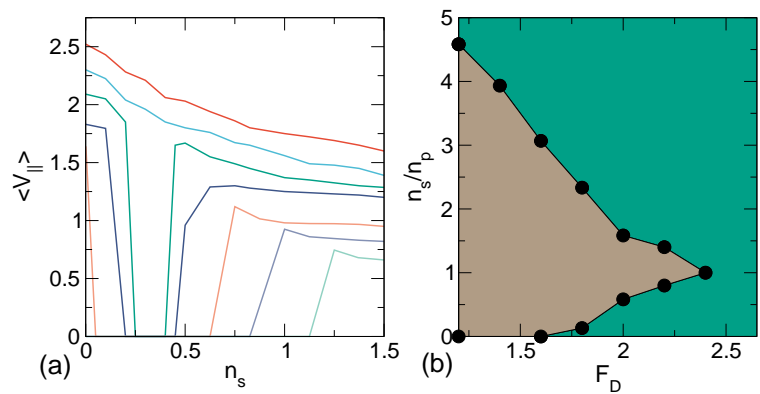}
\end{center}
\caption{
(a) $\langle V_{||}\rangle$ versus skyrmion density $n_{s}$ for the
system in Fig.~\ref{fig:9}(a) with $n_p=0.3$, $F_{p} = 1.6$, and
$\alpha_m/\alpha_d=0.1$
at $F_{D}  = 1.4$, 1.6, 1.8, 2.0, 2.2, 2.4, and $2.6$, from bottom to top.
(b) Dynamic phase diagram for the same system as a function of
$n_{s}/n_{p}$ versus
$F_{D}$.
Green: continuous flow regime; brown: pinned.
}
\label{fig:11}
\end{figure}

In Fig.~\ref{fig:11}(a) we plot $\langle V_{||}\rangle$
versus the skyrmion density $n_{s}$ for the
system in Fig.~\ref{fig:9}(a) with $F_{p} = 1.6$, $n_{p} = 0.3$,
and $\alpha_m/\alpha_d=0.1$ at $F_{D}  = 1.4$, 1.6, 1.8, 2.0, 2.2, 2.4,
and $2.6$.
At $F_{d} = 1.4$ the system is pinned
when $n_{s} \leq 1.0$.
For this skyrmion density, all of the skyrmions can be trapped at
pinning sites and are therefore unable to move since  $F_D<F_p$.
As $n_{s}$ increases,
all of the pinning sites become filled and interstitial skyrmions
appear which are pinned only by repulsion from other skyrmions
directly located at pinning sites. The strength of this
interstitial pinning is
determined by the elastic properties of the skyrmion lattice, and
for these densities it is weaker than $F_p$.
When $F_{D}> F_{p}$, flow can occur even for low $n_s$, where
the driven skyrmion interacts with  
the pinning sites but has few collisions with background skyrmions.
In the limit $n_s=0$ where only the driven skyrmion is present,
the system is always flowing whenever
$F_{D}/F_{p} > 1.0$. 
For the $F_{D} = 2.2$ curve, the
system is flowing up
to $n_{s} = 0.2$ and then 
a pinned region appears
for $0.2 < n_{s} < 0.5$.
At this range of skyrmion densities,
even though
$F_{D} > F_{p}$,
the driven skyrmion experiences a combination of direct pinning from the
pinning sites it encounters plus interstitial pinning by the
nearby directly
pinned skyrmions, giving an additive effect which causes
the apparent pinning strength to be larger than $F_D$.
For $n_{s} > 0.5$, all the pinning sites
start to become occupied and the driven skyrmion experiences only
the weaker interstitial pinning without becoming trapped directly by
any pinning sites.
At small $F_{D}$ it is possible for the driven
skyrmion to become trapped by a pinning site that is already occupied
by a background skyrmion, creating a doubly occupied pinning site,
which is why the value of $n_{s}$ below which the driven skyrmion
can begin to move again shifts to larger $n_s$ with decreasing $F_D$.
The reentrant pinning effect
illustrated in Fig.~\ref{fig:11}(a)
arises from the combination of the direct and interstitial
pinning mechanisms.
In Fig.~\ref{fig:11}(b) we construct a dynamic phase diagram as
a function of $n_{s}/n_{p}$ versus $F_{D}$
for the system in Fig.~\ref{fig:11}(a) showing
the pinned and flowing phases.
Reentrant pinning occurs over the range $F_D=F_p=1.6$ to slightly
above $F_D=2.2$.
The reentrant pinned phase reaches its maximum extent
near $n_{s}/n_{p} = 1.0$, a density at which
the number of directly pinned skyrmions attains its maximum value
while the number of interstitially pinned skyrmions is still nearly zero.

\begin{figure}[h!]
\begin{center}
  \includegraphics[width=10cm]{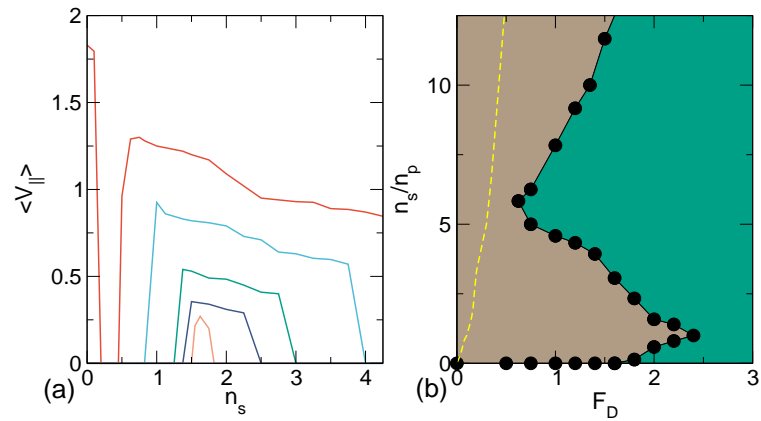}
\end{center}
\caption{ 
(a) $\langle V_{||}\rangle$ versus $n_{s}$ for the system in
Fig.~\ref{fig:11}(a) with $n_p=0.3$, $F_p=1.6$, and $\alpha_m/\alpha_d=0.1$
plotted up to a maximum value of $n_{s} = 4.0$ for         
$F_{D} = 0.5$,
$0.75$,
$1.0$,
$1.2$,
$1.4$,
and $1.6$,
from bottom to top.
(b) Dynamic phase diagram for the same system as
a function of $n_{s}/n_{p}$ versus
$F_{D}$.
Green: continuous flow regime; brown: pinned.
The yellow dashed line indicates the location of the depinning threshold in
systems with no quenched disorder.  
}
\label{fig:12}
\end{figure}

For higher
values of $n_{s}/n_{p}$,
another pinned phase arises
at low values of $F_{D}$
that is produced
by the skyrmion-skyrmion interactions.
In Fig.~\ref{fig:12}(b) we plot $\langle V_{||}\rangle$ versus
$n_{s}$ up to $n_{s} = 4.0$ for 
$F_{D} = 0.5$,
$0.75$,
$1.0$,
$1.2$,
$1.4$,
and $1.6$
in the same system from Fig.~\ref{fig:11}.
At higher $n_{s}$, $\langle V_{||}\rangle$ drops
to zero again
as the system reaches a pinned state.
This second pinned phase is produced by the increase in the
elastic skyrmion-skyrmion interaction energies
at the higher densities.
In the absence of quenched disorder,
the skyrmion-skyrmion interactions are the only mechanism by which
the driven skyrmion can be pinned, and
there is a threshold for motion which increases monotonically 
with increasing $n_{s}$.
When quenched disorder is introduced,
the threshold becomes both non-monotonic and reentrant.
For increasing $F_D$ in Fig.~\ref{fig:12}(a), the elastic energy-induced
pinning transition shifts to higher $n_s$.
In Fig.~\ref{fig:12}(b) we show a
dynamic phase diagram as a function of
$n_{s}/n_{p}$ versus $F_{D}$ for the system in Fig.~\ref{fig:12}(a)
indicating the locations of the pinned and flowing phases.
For $n_{s}/n_{p} < 2.0$, the reentrant pinned state produced by a combination
of direct and interstitial pinning reaches its maximum extent.
As $n_{s}/n_{p}$ increases, the pinned state reaches a minimum width
near $n_{s}/n_{p} = 5.5$, above which the pinned region begins to grow again.
The yellow dashed line
is the depinning threshold in the absence of quenched disorder, which
always falls below the depinning transition of samples with quenched
disorder.
The increase in the depinning threshold due to the addition of pinning
occurs even when the number of skyrmions is significantly larger than
the number of pinning sites since even a relatively small number of pins
can prevent plastic distortions of the background skyrmions, raising the
barrier for motion of the driven skyrmion.

In the phase diagrams of Fig.~\ref{fig:11}(b) and Fig.~\ref{fig:12}(b),
for drives just above the pinned phase,
there is a small window of stick-slip motion (not shown)
which is more prominent for lower values of $n_{s}/n_{p}$.
In addition, within the flowing phase there is 
another critical drive above which
there is an onset of transverse motion, giving 
a finite Hall angle.
This line
has a 
shape similar to that of the depinning curve
but falls at higher values of $F_{D}$.  

In this work we considered a point-like model for skyrmions.
In real skyrmion systems, there is an effective skyrmion
size that can change with field or
exhibit internal modes.
It may be possible that at low fields, the particle picture works well,
while at higher fields,
the skyrmions will start to change shape.
It would be interesting to study
how the effective drag on the driven skyrmion
would change in this case.
Another question regards the distinction between pinning-dominated
pinned states, where direct pinning is responsible for producing the pinning,
and interstitial-dominated or jammed
pinned states, where the pinning of the driven
skyrmion arises from elastic interactions with directly pinned skyrmions.
The fluctuations in the jammed state
generally show that there is a greater tendency for
large scale plastic events to occur, leading to
a larger amount of low frequency noise
compared to the pinning-dominated state.
In work on superconducting vortices with quenched disorder,
the presence of pinned, jammed, and clogged phases
could be deduced by measuring memory effects [\cite{Reichhardt20}].
For the single driven skyrmion, memory could be tested by reversing
the driving direction.
For strong pinning, the trajectory under reversed drive should mirror
that of the forward drive, indicating a memory effect, whereas
in samples with strong plastic distortions,
the trajectory for forward and reversed drive will differ due to the appearance
of plastic distortions in the background skyrmions.
 
\section{Summary}
We have examined the fluctuations and pinning effects for
individually driven skyrmions moving through an assembly of other
skyrmions and quenched disorder.
We find that in the absence of quenched disorder,
there is a depinning force which increases monotonically
with increasing skyrmion density.
When quenched disorder is introduced, the
driven skyrmion experiences a combination of pinning and drag effects
from both the pinning sites and the background skyrmions.
Both with and without quenched disorder, there is a second, higher driving
threshold for the onset of motion transverse to the drive and the appearance
of a finite skyrmion Hall angle.
For higher drives, addition of quenched disorder actually
increases the velocity of the driven skyrmion since the pinning sites
help prevent the background skyrmions from being dragged along by the
driven skyrmion.
Near depinning, in the absence of quenched disorder
the velocity fluctuations show a combination
of periodic oscillations from the elasticity of the ordered background
skyrmion lattice
along with stronger jumps associated with plastic distortions of
the background skyrmions.
This produces a velocity power spectrum that has
narrow band noise peaks superimposed on
a $1/f^{\alpha}$ shape with $\alpha = 1.2$.
As the drive increases,
the spectrum becomes white,
and for very high drives,
a strong narrow band signature emerges once
the driven skyrmion is moving too rapidly to
generate plastic distortions in the background skyrmions.
Addition of quenched disorder
reduces the
frequency of
plastic events, giving a white noise spectrum.
In the absence of disorder, a damping-dominated system
generally shows strong narrow band noise fluctuations
as the driven skyrmion moves along one-dimensional paths in
the background skyrmion lattice, whereas
in the Magnus-dominated limit,
the driven skyrmion moves at an angle through the lattice,
generating
dislocations
and reducing the strength of the narrow band signature.
When the disorder is strong, the driven skyrmion can undergo
stick-slip motion
due to a combination of being trapped at pinning sites and
interacting elastically with the background skyrmions, which
produces a bimodal velocity distribution along with $1/f^{\alpha}$ noise. 
For systems with quenched disorder,
the depinning threshold is
highly non-monotonic as a function of the skyrmion density, passing through
both peaks and minima.
This is due to a competition between two different pinning effects.
The depinning threshold drops when the number of skyrmions becomes
larger than the number of pinning sites since the driven skyrmion must be
pinned through interstitial interactions with directly pinned skyrmions
instead of sitting in a pinning site directly;
however,
at higher densities, the increasing strength of the elastic interactions
between the skyrmions causes the depinning threshold to rise again with
increasing density.
At low densities the system can be viewed as being in a pinning-dominated
regime, while
at higher densities it is in an interstitial-dominated or
jamming regime.   
Beyond skyrmions, our results should be relevant to fluctuations
in other particle-based systems
such as individually dragged vortices in type-II superconductors.

\section*{Conflict of Interest Statement}

The authors declare that the research was conducted in the absence of any commercial or financial relationships that could be construed as a potential conflict of interest.

\section*{Author Contributions}

All authors contributed equally to all portions of this work.

\section*{Funding}
We gratefully acknowledge the support of the U.S. Department of
Energy through the LANL/LDRD program for this work.
This work was supported by the US Department of Energy through
the Los Alamos National Laboratory.  Los Alamos National Laboratory is
operated by Triad National Security, LLC, for the National Nuclear Security
Administration of the U. S. Department of Energy (Contract No. 892333218NCA000001).

\bibliographystyle{frontiersinSCNS_ENG_HUMS}
\bibliography{mybib}

\section*{Figure captions}

\end{document}